\documentclass[prb,twocolumn,floatfix,showpacs]{revtex4}
\usepackage{graphicx}
\newcommand{\be}{\begin{equation}}
\newcommand{\ee}{\end{equation}}
\newcommand{\bea}{\begin{eqnarray}}
\newcommand{\eea}{\end{eqnarray}}
\newcommand{\bm}[1]{\mbox{\boldmath$#1$}}
\newcommand{\ra}{\rangle}
\newcommand{\la}{\langle}
\begin{document}
\title{Scattering functions of knotted ring polymers}

\author{Miyuki K. Shimamura}
\email{miyuki@degway.phys.ocha.ac.jp}
\affiliation{Graduate School of Humanities and Sciences, Ochanomizu
University, 2-1-1 Ohtsuka, Bunkyo-ku, Tokyo 112-8610, Japan}
\author{Kumiko Kamata}
\email{kamakama@degway.phys.ocha.ac.jp}
\affiliation{Graduate School of Humanities and Sciences, Ochanomizu
University, 2-1-1 Ohtsuka, Bunkyo-ku, Tokyo 112-8610, Japan}
\author{Akihisa Yao}
\email{yao@degway.phys.ocha.ac.jp}
\affiliation{Graduate School of Humanities and Sciences, Ochanomizu
University, 2-1-1 Ohtsuka, Bunkyo-ku, Tokyo 112-8610, Japan}
\author{Tetsuo Deguchi}
\email{deguchi@phys.ocha.ac.jp}
\affiliation{Department of Physics, Ochanomizu University, 2-1-1
Ohtsuka, Bunkyo-ku, Tokyo 112-8610, Japan}

\date{\today}

\begin{abstract}
We discuss the scattering function of a Gaussian random polygon 
with $N$ nodes under a given topological constraint 
through simulation. 
We obtain  the Kratky plot of a Gaussian polygon of $N=200$ 
having a fixed knot for some different knots such as 
the trivial, trefoil and figure-eight knots. 
We find that some characteristic properties 
of the different Kratky plots are consistent with   
the distinct values of the mean square radius 
of gyration for Gaussian polygons with the different knots.      
\end{abstract}

\pacs{82.35.Lr,05.40.Fb,05.20.-y}

%
%

\maketitle

\section{\label{sec:intro}Introduction}

Ring polymers have attracted much interest in polymer physics, 
and various properties have been studied both theoretically 
and experimentally. 
\cite{Casassa,Vologodskii,Roovers,Semlyen,Burchard,tenBrinke} 
 For the Gaussian random polygon 
the analytic expression of the static structure factor  
was obtained by Casassa. \cite{Casassa} 
The Kratky plot of the scattering function is compared  
with that of star polymers with four or five arms. \cite{Casassa,Burchard}
The scattering data of cyclic polystyrene 
in deuteriated cyclohexane 
was obtained by the SANS experiment. 
 \cite{tenBrinke}

Recently statistical properties of ring polymers 
under  topological constraints have been  investigated extensively mainly 
through computer simulation. \cite{desCloizeaux-Let,Deutsch,GrosbergPRL,Miyuki,Akos,Matsuda,Moore,PRE2001R,PRE-Miyuki}   
It is first conjectured by des Cloizeaux that a topological constraint 
should lead to  
an effective repulsion among segments of ring polymers \cite{desCloizeaux-Let}. 
The conjecture is supported by the numerical observations that  
 the mean square  radius of gyration  of ring polymers  with a fixed knot 
is larger than that of no topological constraint. \cite{Deutsch,Miyuki,Akos,Matsuda,Moore,PRE2001R,PRE-Miyuki} 
In fact, the topological swelling is observed 
particularly for ring polymers 
with small or zero  excluded volume. \cite{PRE2001R,PRE-Miyuki}

In this paper we discuss the scattering function of 
ring polymers under a topological constraint. It should be fundamental   
for studying ring polymers in scattering  experiments. 
We consider ring polymers in solution at the $\theta$ temperature, 
and they are modeled by random polygons. Here, random polygons have   
 no excluded volume, i.e. they have no thickness.  
We shall evaluate the radial distribution function in simulation, 
and then take the Fourier transformation. 
We shall show  the Kratky plot of the scattering function of a random polygon  
having some fixed knot type.

\section{\label{sec:method}Simulation Methods}

Making use of the conditional probability distribution 
\cite{desCloizeaux-Mehta}, 
we have systematically constructed $10^5$ samples 
of the Gaussian random polygon with $200$ nodes. 
We have calculated two knot invariants,  $\Delta_K(-1)$ and $v_2(K)$, 
 to each of the $10^5$ configurations, and effectively classified 
them into different topological classes. 
Here the symbol  $\Delta_K(-1)$ denotes the determinant of a knot $K$, 
which is given by the Alexander polynomial $\Delta(t)$ 
evaluated at $t=-1$.  The symbol $v_2(K)$ 
is the  Vassiliev invariant of the second degree. 
We select such polygons that have the same set 
of values of the two knot invariants.  
Here we calculate $v_2(K)$ by the algorithm \cite{Polyak}.
The two knot invariants are 
practically useful for computer simulation 
of random polygons with a large number of polygonal nodes \cite{DeguchiPLA}.

We consider four different topological classes: 
the trivial knot ($0$), the trefoil knot ($3_1$), 
the figure-eight knot  ($4_1$) and the other knots ($other$), in the paper.     
We denote by ``$all$'' such polygons that have no topological constraint.

 Let us denote by $\la R_{G,K}^2 \ra$ 
the mean square radius of gyration for such  
Gaussian polygons that have $N$ nodes 
and a given topological constraint $K$.   
The estimates of  $\la R_{G,K}^2 \ra$ for $N=200$ 
are given in Table \ref{tab:gyr}  for several knots. 
For different numbers of $N$, they have been obtained in Ref. 
 \onlinecite{Miyuki}. The fraction of Gaussian polygons with  
a given knot has also been evaluated for some knots. \cite{Tsurusaki}  
 
\begin{table}[th]
{\begin{tabular}{ccc}
\hline 
$K$ & $\la R_{G,K}^2 \ra$ & errors  \\
\hline 
 $0$  & 18.033  &     0.082 \\ 
 $3_1$   & 16.208  &    0.120  \\  
 $4_1$  & 15.043  &    0.250  \\  
 $other$ &  13.459  &    0.118 \\   
$all$ & 16.674  &   0.060  \\
\hline 
\end{tabular}} 
\caption{\label{tab:gyr} Mean square radius of gyration $\la R_{G,K}^2 \ra$ 
of a Gaussian random polygon with a topological condition $K$. Here $N=200$. }  
\end{table}

\begin{figure}
\includegraphics[width=8cm]{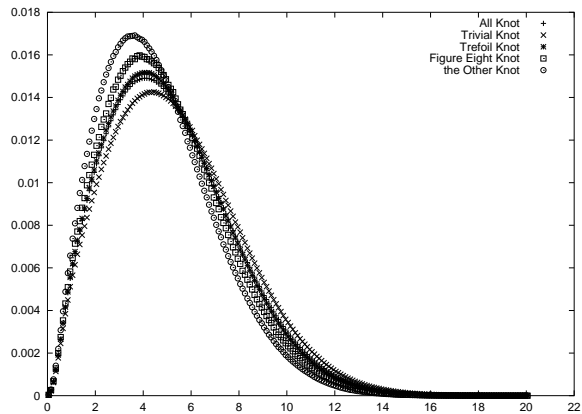} 
\caption{\label{fig:paircorr} 
The probability distribution 
$4 \pi r^2 \, g_K(r) \Delta r/N$
versus the distance $r$.    
Here  $N=200$ and $\Delta r = 0.1$. 
The plots of the five topological conditions, 
$all$, $0$, $3_1$, $4_1$ and $other$,  are represented by crosses, 
tilted crosses, double crosses, open squares and open circles, respectively.    }
\end{figure}

\section{Radial distribution functions}

Let us define the segment pair correlation function 
of a random polygon with $N$ nodes by \cite{Doi} 
\be 
g({\bm r}) = {\frac 1 N} \sum_{m,n=1}^{N} \la 
\delta({\bm r} - ({\bm R}_m- {\bm R}_n)) \ra  \, . 
\ee
Here ${\bm R}_m$ denotes the position vector of the $m$th node 
for $m=1, 2, \ldots, N$.  
Due to spherical symmetry, the function 
$g({\bm r})$ depends only on the distance, $r=|{\bm r}|$, 
and we denote it  by $g(r)$. We call it    
  the radial distribution function. 
Here we note that $4 \pi r^2 g(r) \Delta r/N$ gives 
the probability of other segments appearing in a spherical shell
from radius $r$ to $r+ \Delta r$ centered at a given segment.

 For a Gaussian polygon under a topological condition $K$, 
we denote by $g_K({\bm r})$ and $g_K(r)$ 
 the pair correlation function and the radial distribution function, 
respectively. 
 The graphs of the probability distribution 
$4 \pi r^2 g_K(r) \Delta r /N $ are plotted  against $r$ 
in Fig. \ref{fig:paircorr}. 
They are consistent with a preliminary result \cite{proceedings}.

We now generalize the recent observation 
that for random polygons 
the peak position of the probability distribution  
$4 \pi r^2 g(r) \Delta/N$ should be given by the gyration radius. 
\cite{Kamata}   
Let us denote by $R_{G,K}$ 
the square root of the mean-square radius of gyration, 
i.e. $R_{G,K} = \sqrt{\la R_{G,K}^2 \ra}$.  
The estimates of the peak position of the probability distribution 
$4 \pi r^2 g_K(r) \Delta r /N $ for the five topological conditions are 
listed in Table \ref{tab:peak} together with those 
of the gyration radius, $R_{G,K}$.   
They are almost identical up to numerical errors. Thus, we have 
$r_{\rm peak}= R_{G,K}$ within errors. This is characteristic to ring polymers. 
In fact, for a 
Gaussian linear chain, the peak position of the probability 
distribution $4 \pi r^2 g(r) \Delta r/N$ is located at  
$r \approx 0.74 R_{G,lin}$, where $R_{G,lin}$ denotes the square root of 
the mean square radius of gyration for the linear chain. \cite{Teraoka} 

\begin{table}[th]
{\begin{tabular}{ccc}
\hline 
$K$ &  $r_{\rm peak}$  & $R_{G,K}$  \\
\hline 
 $0$  &  4.25  &  4.246 \\ 
 $3_1$   & 4.05  &    4.026  \\  
 $4_1$  & 3.85  &    3.879  \\  
 $other$ &  3.65  &  3.669 \\   
$all$ & 4.05  &  4.083 \\
\hline 
\end{tabular}} 
\caption{ \label{tab:peak} 
Peak position $r_{\rm peak}$ of the probability distribution
 $4 \pi r^2 g_K(r) \Delta r/N$ for a  topological condition $K$. 
The  estimates of $r_{\rm peak}$ may have errors of order 0.1 at most. }  
\end{table}

\section{\label{sec:scattering}Scattering functions }

For a random polygon with $N$ nodes,  we define 
the (single-chain) static structure factor $g({\bm q})$ by 
the Fourier transform of the pair correlation function as follows \cite{Doi}
\be 
g({\bm q}) = \int d{\bm r} e^{i {\bm q} \cdot {\bm r}} g({\bm r})  
= {\frac 1 N} \sum_{m,n=1}^{N} 
\la \exp( i {\bm q} \cdot ( {\bm R}_m - {\bm R}_n)) \ra 
\label{eq:g(q)}
\ee
We also call it the scattering function. 
The scattering function $g({\bm q})$ 
depends only on $q=|{\bm q}|$, and we denote it by $g(q)$.  
 For a Gaussian polygon under a topological condition $K$, 
we denote the static structure factor or 
the scattering function  by the symbol $g_K(q)$.

\begin{figure}
\includegraphics[width=8cm]{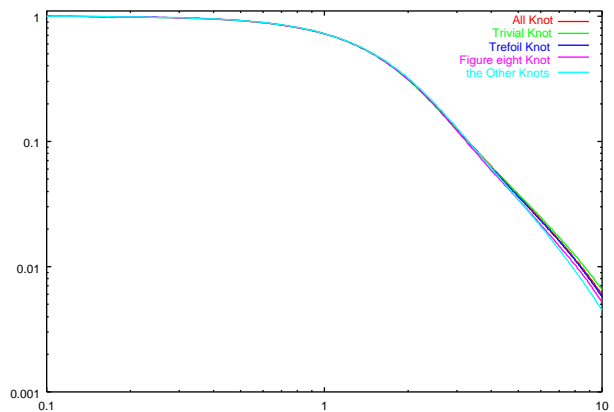}
\caption{\label{fig:dblog} Double logarithmic plot of 
the form factor $P_K(q)$ of an $N$-noded  Gaussian polygon 
under a topological condition $K$ versus 
the variable $u=q R_{G,K}$. The curves colored with 
red, green, blue, magenta, and cyan correspond to the 
cases of $all$, $0$, $3_1$, $4_1$ and $other$, respectively. }
\end{figure}

Let us introduce the form factor $P_K(q)$ 
for a Gaussian polygon under a topological condition $K$ 
as follows 
\be 
P_K(q) = {\frac {g_K(q)} {g_K(0)}}
\ee
We have $P_K(q)= g_K(q)/N$ from (\ref{eq:g(q)}). 
We have evaluated the form factor $P_K(q)$ 
for the five topological conditions. 
Let us define variable $u$ by $u=q R_{G,K}$.  
The double logarithmic plot of 
the form factor $P_K(q)$ versus $u$ 
is shown in  Fig. \ref{fig:dblog}. Here we recall that the form factor 
$P_{all}(q)$ was evaluated analytically in terms 
of the Dawson integral. \cite{Casassa}

In the region  from $u=0$ up to $u=2$ or 3, 
the form factors $P_K(q)$ for the five topological conditions  
overlap each other. 
For $u > 5$, 
the graphs of the different topological conditions  
make parallel lines.  The gradient is almost given by $-2$, 
which is consistent with 
the Gaussian asymptotic behavior.

\begin{figure} 
\includegraphics[width=9cm]{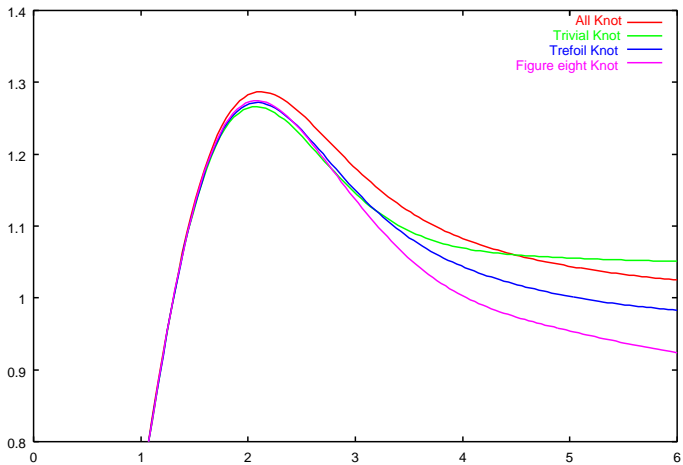} 
\caption{\label{fig:Kratky} Plot of $(q R_{G,K})^2 \, P_K(q)$
versus  $u=q R_{G,K}$. The Kratky plots of 
the trivial, trefoil and figure-eight knots are represented by 
the green, blue and magenta curves, respectively. The red curve corresponds to 
the case of no topological constraint.  
}
\end{figure}

Let us discuss the Kratky plots of the form factors $P_K(q)$ for 
some different topological conditions. 
The plots of $(q R_{G,K})^2 P_K(q)$ versus the variable 
$u=q R_{G,K}$ are shown in Fig. \ref{fig:Kratky}.    
Here we have numerically evaluated the Fourier transformations of the 
radial distribution functions  
by an interpolation method \cite{Kamata}.

\begin{center} 
\begin{table}[th]
{\begin{tabular}{ccc}
\hline 
$K$ &  $u_{\rm peak}$  & peak height  \\
\hline 
 $0$  &  2.08 & 1.266  \\ 
 $3_1$   & 2.09  &    1.272  \\  
 $4_1$  & 2.09  &    1.275  \\  
 $other$ &  2.09  &  1.300  \\   
$all$ & 2.08 &  1.287  \\
\hline 
\end{tabular}} 
\caption{ \label{tab:u-peak} 
Peak position $u_{\rm peak}$ of the Kratky plot 
 $(q R_{G,K})^2 \, P_K(q)$ versus $u=q R_{G,K}$ 
for the Gaussian random polygon under 
a given topological condition $K$. 
The $q$ value is given by an integral multiple of $0.01$. }  
\end{table}
\end{center}

For the graphs shown in Fig. \ref{fig:Kratky} 
we observe that the mean square radius of gyration, $\la R_{G,K}^2 \ra$,  
plays a central role 
in the scattering function of the Gaussian polygon 
under a topological condition, $K$.  
In the small $u$ region such as $u < 2$, the plots for the different   
topological conditions overlap completely, while for $u > 2$ the 
graphs become separate. Here, the peak positions of the Kratky plots 
are given by the same value of $u$ for all the five topological conditions. 
The estimates of the peak positions are given in Table \ref{tab:u-peak}. 
The peak height depends on topological conditions. 
The Kratky plot for the trivial knot has the smallest peak height.
The peak height for the trefoil knot is a little larger than that 
of the trivial knot. 
However, for the Kratky plots of the trefoil and figure-eight knots, 
the peak heights are given by almost the same value.  

The Kratky plots of  Fig. \ref{fig:Kratky} are not in contradiction 
with those of previous studies, and even generalize them.    
For lattice random polygons with $N=160$, the Kratky plots 
were numerically evaluated for all polygons and knotted polygons, 
respectively.  \cite{tenBrinke}
Since the majority of polygons with nontrivial knots for $N=200$ 
are given by those of the trefoil knot, the Kratky plots of 
 all polygons and knotted polygons in Ref. \onlinecite{tenBrinke} 
approximately correspond to 
those of no topological constraint and the trefoil knot 
shown in Fig. \ref{fig:Kratky}, respectively.

For $u > 5$,  we observe 
that each of the Kratky plots of Fig. \ref{fig:Kratky} 
approach constant values. 
We thus suggest an asymptotic behavior that for any topological constraint $K$ 
the form factor $P_K(q)$ should become close to 
that of the Gaussian polygon for $u \gg 1$, such as  
$P_K(q) \propto 1/(q R_{G, all})^2$. 
Here we recall that when $u \gg 1$ the form factor $P_{lin}(q)$ of a 
Gaussian linear chain is  
approximated by $P_{lin}(q) \approx 2/(q^2 {R^2_{G,lin}})$. 
We thus have  
\be
(q R_{G,K})^2 \, P_K(q) \propto {\frac {R_{G, K}^2} {R_{G, all}^2}} 
 \quad (u \gg 1) \label{eq:asympt}
\ee 
The asymptotic constant value of the Kratky plot of a knot $K$ should 
become smaller when the knot $K$ becomes more complex.  
It is consistent with the observation of Fig. \ref{fig:Kratky} that   
the order of the asymptotic constant values are the same with that of 
the values $\la R_{G, K}^2 \ra$ in Table \ref{tab:gyr}, for the three knots. 
For $u > 10 $, systematic errors of the Kratky plots 
may be larger than statistical errors of the gyration radius  $R_{G,K}$.

The distribution function between two given nodes 
of a random polygon with a fixed knot 
is recently evaluated through simulation, and 
it is found to be close to the Gaussian one. \cite{Yao} 
The interpretation (\ref{eq:asympt}) should be  also  consistent 
with the observation.

\section{\label{sec:conclusion}Conclusion}

We have evaluated the scattering functions of Gaussian 
random polygons with $N=200$ under different topological conditions $K$. 
The Kratky plots have been obtained 
for the different $K$. 
They overlap up to $u=2$, and they become separate for $u > 2$, 
and they approach constant values for $u \gg 1$.     
Several characteristic properties are explained in terms 
of the different values  of the gyration radius $R_{G,K}$.


\begin{thebibliography}{99}

\bibitem{Casassa} E.~F. Casassa, J. Polym. Sci., Part A {\bf 3},  
605 (1965). 

\bibitem{Vologodskii} A.V. Vologodskii, A.V. Lukashin, M.D.
Frank-Kamenetskii, and V.V. Anshelevich,
Sov. Phys. JETP {\bf 39}, 1059 (1974).

\bibitem{Roovers} J.~R. Roovers and P.~M. Toporowski, 
Macromolecules {\bf 16}, 843 (1983).  

\bibitem{Semlyen}  
{\it Cyclic Polymers}, ed. J.A. Semlyen 
(Elsevier Applied Science Publishers, London and New York, 1986);  
2nd Edition (Kluwer Academic Publ., Dordrecht, 2000). 
  

\bibitem{Burchard} W. Burchard,  
in {\it Cyclic Polymers}, ed. J.A. Semlyen 
(Elsevier Applied Science Publishers, London and New York, 1986) pp. 43--84.   

\bibitem{tenBrinke} G. ten Brinke and G. Hadziioannou, 
Macromolecules {\bf 20}, 480 (1987).  


\bibitem{desCloizeaux-Let} J. des Cloizeaux, 
J. Physique Letters (France) {\bf 42} , L433 (1981).

\bibitem{Deutsch}
J.~M. Deutsch, Phys. Rev. E {\bf 59}, R2539 (1999).

\bibitem{GrosbergPRL}
A.~Yu. Grosberg, Phys. Rev. Lett. {\bf 85}, 3858 (2000).


\bibitem{Miyuki} M.~K. Shimamura and T. Deguchi, 
J. Phys. A: Math. Gen. {\bf 35}, L241 (2002).

\bibitem{Akos} A. Dobay, J. Dubochet, K. Millett, P.~E. Sottas and 
A. Stasiak, Proc. Natl. Acad. Sci. USA {\bf 100}, 5611 (2003).   

\bibitem{Matsuda} H. Matsuda, A. Yao, H. Tsukahara, T. Deguchi, 
K. Furuta and T. Inami,   
Phys. Rev. E {\bf 68}, 011102 (2003). 

\bibitem{Moore} N.~T. Moore, R.~C. Lua and A.~Y. Grosberg, 
 Proc. Natl. Acad. Sci. USA {\bf 101}, 13431 (2004).


\bibitem{PRE2001R} M.~K. Shimamura, and T. Deguchi, 
Phys. Rev. E {\bf 64}, 020801(R) (2001) 


\bibitem{PRE-Miyuki} M.~K. Shimamura and T. Deguchi, 
Phys. Rev. E {\bf 65}, 051802 (2002).



\bibitem{desCloizeaux-Mehta} J.~ des Cloizeaux and M.~L. Mehta, 
J. Phys. (Paris) $\bf{40}$, 665 (1979). 


\bibitem{Polyak} M. Polyak and O. Viro, 
Int. Math. Res. Not. No.11, 445 (1994).

\bibitem{DeguchiPLA}
T. Deguchi and K. Tsurusaki, Phys. Lett. A {\bf 174}, 29 (1993).

 
\bibitem{Tsurusaki} T. Deguchi and K. Tsurusaki,  
Phys. Rev. E. {\bf 55}, 6245 (1997) . 


\bibitem{Doi} M. Doi, {\it Introduction to Polymer Physics} (Clarendon Press, 
Oxford, 1996) 


\bibitem{proceedings} M.~K. Shimamura and T. Deguchi, 
preprint.  



\bibitem{Teraoka} I. Teraoka, {\it Polymer Solutions}  
(John Wiley $\&$ Sons, Inc., New York, 2002). 


\bibitem{Kamata} K. Kamata, Scattering functions of 
circular polymers (in Japanese),  
Master thesis,  Ochanomizu University, March 2005. 


\bibitem{Yao} A. Yao, H. Tsukahara, T. Deguchi and T. Inami, 
J. Phys. A: Math. Gen. {\bf 37}, 7993 (2004). 


\end{thebibliography}
\end{document}